# Magnetic Particle Spectroscopy: A Short Review of Applications

Kai Wu[†,*], Diqing Su[‡], Renata Saha[†], Jinming Liu[†], Vinit Kumar Chugh[†], and Jian-Ping Wang[†,*]

[†]Department of Electrical and Computer Engineering, University of Minnesota, Minneapolis, Minnesota 55455, USA

[‡]Department of Chemical Engineering and Material Science, University of Minnesota, Minneapolis, Minnesota 55455, USA

*E-Mails: wuxx0803@umn.edu (K.W.), jpwang@umn.edu (J.-P.W.).

**Abstract**

Magnetic particle spectroscopy (MPS), also called magnetization response spectroscopy, is a novel measurement tool derived from magnetic particle imaging (MPI). It can be interpreted as a zero-dimensional version of MPI scanner. MPS was primarily designed for characterizing superparamagnetic iron oxide nanoparticles (SPIONs) regarding their applicability for MPI. In recent years, it has evolved into an independent, versatile, highly sensitive, inexpensive platform for biological and biomedical assays, cell labeling and tracking, and blood analysis. MPS has also developed into an auxiliary tool for magnetic imaging and hyperthermia by providing high spatial and temporal mappings of temperature and viscosity. Furthermore, other MPS-based applications are being explored such as magnetic fingerprints for product tracking and identification in supply chains. There are a variety of novel MPS-based applications being reported and demonstrated by many groups. In this short review, we highlighted some of the representative applications based on MPS platform, thereby providing a roadmap of this technology and our insights for researchers in this area.

## 1. Introduction

Magnetic particle spectroscopy (MPS) is a flourishing research area that closely relates to magnetic particle imaging (MPI). While MPI directly measures and maps the concentration of superparamagnetic iron oxide nanoparticle (SPION) over a spatial position, MPS is interpreted as a zero-dimensional MPI scanner which conducts spectroscopic studies on SPIONs. Although high-moment magnetic nanoparticles (MNPs) such as iron nanoparticles can provide higher magnetic signal than SPIONs of same size and enhance the sensitivity, the biocompatibility of these types of high-moment MNPs need to be further investigated.[1–7] Herein, in this work we only focus on the MPS-based applications using SPIONs. Primarily the MPS platform was dedicated to assessing the performance of the SPION magnetic tracers regarding their applicability for MPI.[8–13] Over the years, it has developed into a highly sensitive, fast and versatile sensing platform for a wide variety of biological and biomedical assays.[14–18] In MPS, sinusoidal magnetic fields are applied to periodically magnetized SPIONs. The magnetic moments of SPIONs tend to align with the applied fields through relaxation processes (Néel and Brownian processes), which are countered by the thermal fluctuation. In addition, the relaxation processes are directly linked to the physical conditions such as the viscosity and temperature of SPION aqueous medium as well as the bound states of target analytes (chemicals/biological compounds) to the SPIONs.[16,17,19–29] On the other hand, SPIONs, with their size comparable to chemical/biological compounds, have been the subject of increasing interest in the areas of medical diagnostics and therapy. In general, a SPION consists of a magnetic core (magnetite/maghemite) coated with inorganic and/or organic capping layers (e.g., dextran, chitosan, silica, etc.) to prevent aggregation and to improve physicochemical stability under various physiological environments, as shown in Figure 1(A). In addition, suitable surface functionalization of ligands/antibodies/aptamers/proteins onto SPIONs allows for highly selective chemical interactions with biological systems. Due to the negligible magnetic background from biological samples, highly sensitive detection could be achieved by using SPIONs as contrasts/markers/tracers.

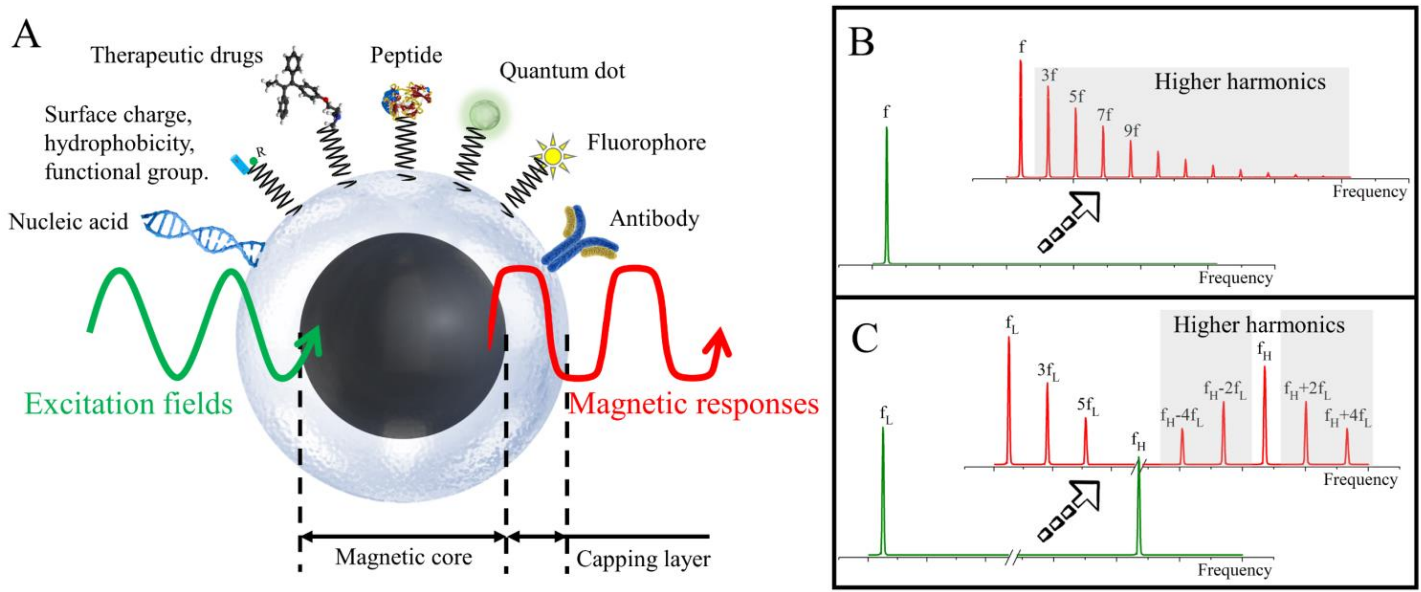

Figure 1. (A) Schematic drawing of a multi-functional SPION. Higher odd harmonics are generated by the dynamic magnetic responses of SPIONs under external excitation magnetic fields. (B) shows the spectra of mono-frequency excitation field and the resulting MPS spectra. (C) shows the spectra of dual-frequency excitation fields and the resulting MPS spectra.

In MPS, the dynamic magnetic responses of SPIONs, which contain unique higher odd harmonics, are harvested by a pair of pick-up coils. The MPS spectra (higher harmonic amplitude and phase) is obtained by Fourier transform of the detected time domain signal, as shown in Figure 1(B) & (C). Since the harmonic amplitude is largely dependent on the quantity of SPIONs in the testing sample, the MPS results could be biased by the deviations of SPION quantities in each sample, especially for the scenarios of detecting very low abundancy of target analytes.[14] Due to this concern, other SPION quantity-independent metrics have been reported and demonstrated for bioassays, such as magnetic susceptibility, harmonic phase angle (or phase lag), and harmonic amplitude ratios (the 3rd over the 5th harmonic ratio R35 or the 5th over the 3rd harmonic ratio R53).[17,29–34] In recent years, MPS has shown a strong capability of sensing the subtle changes in any property influencing the Néel and/or Brownian processes of the SPIONs within seconds.

MPS bears the superior advantages over conventional magnetometers by providing fast and simple measurement procedures for SPION systems without the need of cooling the device. The working mechanisms and theories of MPS platform have been extensively reported by several papers.[8,14,18,35] Herein, in this review, we focus on the MPS derived applications. We firstly reviewed the MPS-based application for biological and biomedical assays using surface functionalized SPIONs as magnetic markers, followed by MPS-based cell analysis where the SPIONs are used as contrast agents to monitor cell uptake, passage, and cell vitality.[17,29,36–39] With superior biocompatibility, biodegradability and superparamagnetic nature, SPIONs are much desired magnetic markers/tracers for the aforementioned applications. In addition, MPS has also been exploited as an auxiliary tool for applications such as magnetic imaging and hyperthermia by providing real-time viscosity and temperature mapping, as well as mechanical force sensor in ball mills.[15,21,40,41] Furthermore, other MPS-based applications that are sometimes overlooked but owns great market potentials such as the characterization of SPIONs from aqueous medium, the marking/tracking of products in the complex global resource and supply chains, the determination of total blood volume and the evaluation of blood clot progression.[15,16,42–46] In this short review, we are only able to list some of the representative applications and the MPS-based applications are not limited to this paper. We hope this review could provide insights into the great potential of MPS and serve as a roadmap for researchers in this area.

## 2. MPS-based Biological and Biomedical Assays
### 2.1 MPS-based Immunoassays

As is aforementioned, the magnetic moments of SPIONs relax to align with the external magnetic field through the joint Néel and Brownian processes. Néel process is the internal flipping of magnetic moment inside a stational SPION while Brownian process is the physical rotation of the entire SPION along with its magnetic moment. While both processes are countered by the thermal fluctuation, Néel process is also affected by the crystal anisotropy and volume of the magnetic core. On the other hand, Brownian process is affected by the liquid medium viscosity and hydrodynamic volume of the SPION (volume of magnetic core and surface attachments). For free and non-interacting SPIONs with magnetic core size below 20 nm, Néel process guides the dynamic magnetic responses of SPIONs, while Brownian process becomes the dominating factor when magnetic core size of SPION is above 25 nm.[42,47] For the applications of MPS-based immunoassays, Brownian process-dominated SPIONs are most sensitive to the binding events of target analytes, thus, are frequently used for this purpose.

Herein, a streptavidin and biotin system with high binding affinity is at first demonstrated as the model system, shown in Figure 2(A). Streptavidin is a crystalline tetrameric protein with a molecular weight of $4 \times 15$ kDa. SPIONs are surface functionalized with biotins. Each streptavidin molecule can bind up to 4 biotins with high binding affinity, which allows the interaction with multiple biotin moieties from different SPIONs, and, as a result, forms SPION clusters. Six SPION samples are prepared and each sample is added with streptavidin of different concentrations. The harmonic ratio R35 is plotted as a function of driving field frequency, which increases as the concentration/quantity of streptavidin increases, as shown in Figure 2(B). Transmission electron microscopy (TEM) images are taken to investigate the SPION clusters in six samples. Well dispersed SPIONs are observed from the negative control group (0 nM streptavidin), whereas bigger SPION clusters are observed from TEM images as the concentration of streptavidin increases, as shown in Figure 2(C). In addition, different bound state models are presented. Due to the fact that each tetrameric streptavidin hosts up to four biotin binding sites, SPIONs could form clusters, chains, tetramers, trimers, dimers, and so forth.

Zhang *et. al* reported MPS-based bioassay using an aptamer-thrombin system.[17] Two anti-thrombin aptamers (15 mer and 29 mer) that can specifically bind to two different epitopes from thrombin are selected. Two groups of SPIONs are prepared, each group is surface functionalized with one type of anti-thrombin aptamer, as shown in Figure 2(D). Each thrombin hosts two hetero-binding domains that allows two SPIONs with different aptamers to bind with. The harmonic ratio R53 is collected as SPION quantity-independent metric. Results show that adding high concentration thrombin to either the 15 mer or 29 mer aptamer functionalized SPIONs alone produce very limited change in R35 metric. Larger R35 changes are observed in the presence of both groups of SPIONss, when 15 mer and 29 mer aptamers bind to hetero-binding domains from thrombin, resulting in SPION clusters. Similarly, Wu *et. al* reported MPS-based bioassay using a polyclonal antibody-antigen system.[48] In order to artificially induce SPION clusters, IgG polyclonal antibodies are anchored onto SPION surfaces. The target analyte, H1N1 nucleoprotein (NP) hosts multiple different epitopes for these IgG polyclonal antibodies. In the presence of H1N1 NP, cross-linking takes place between SPIONs (IgG from particle surface) and H1N1 NP, which forms SPION clusters, as shown in Figure 2(E). The harmonic ratio R35 is collected, and this system is reported to detect as low as 44 nM (4.4 pmole) H1N1 nucleoprotein.

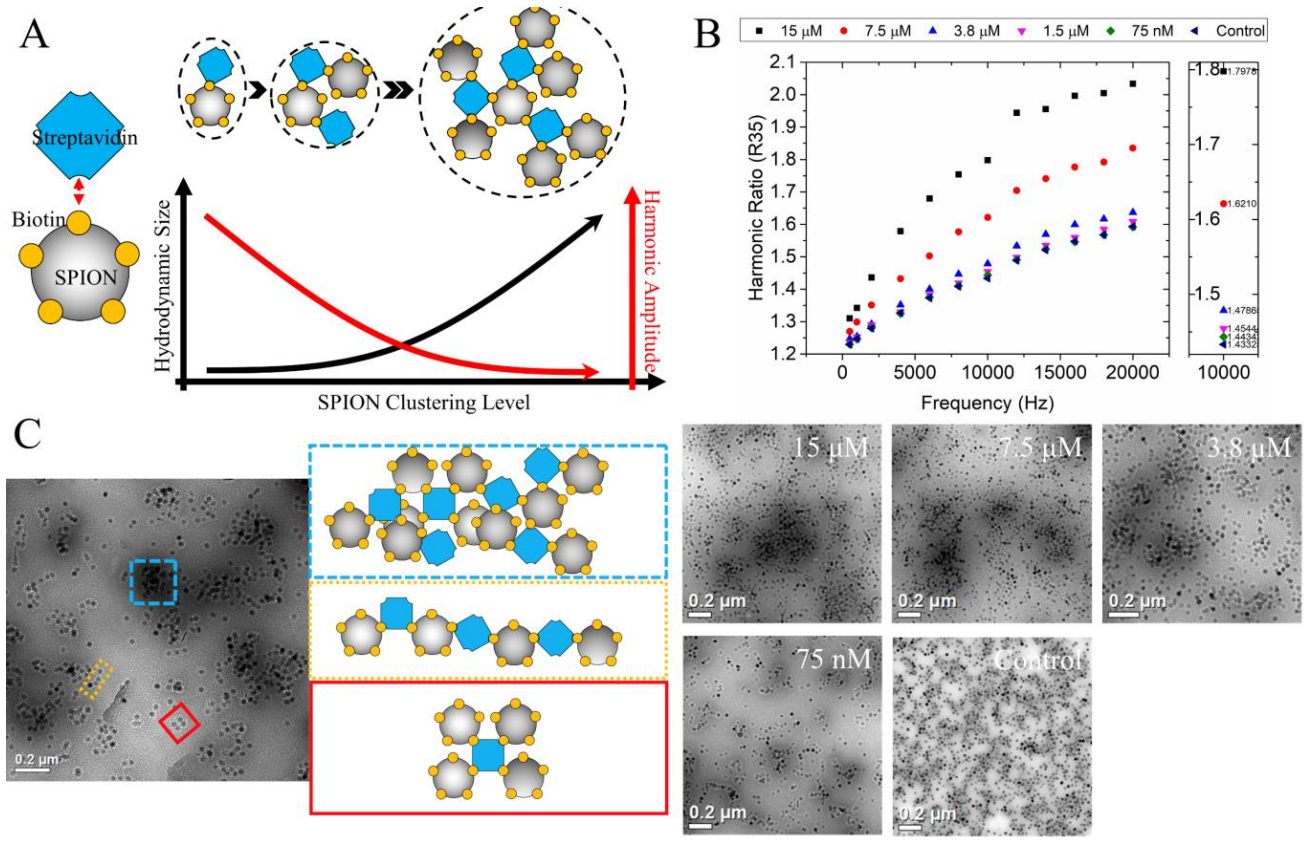
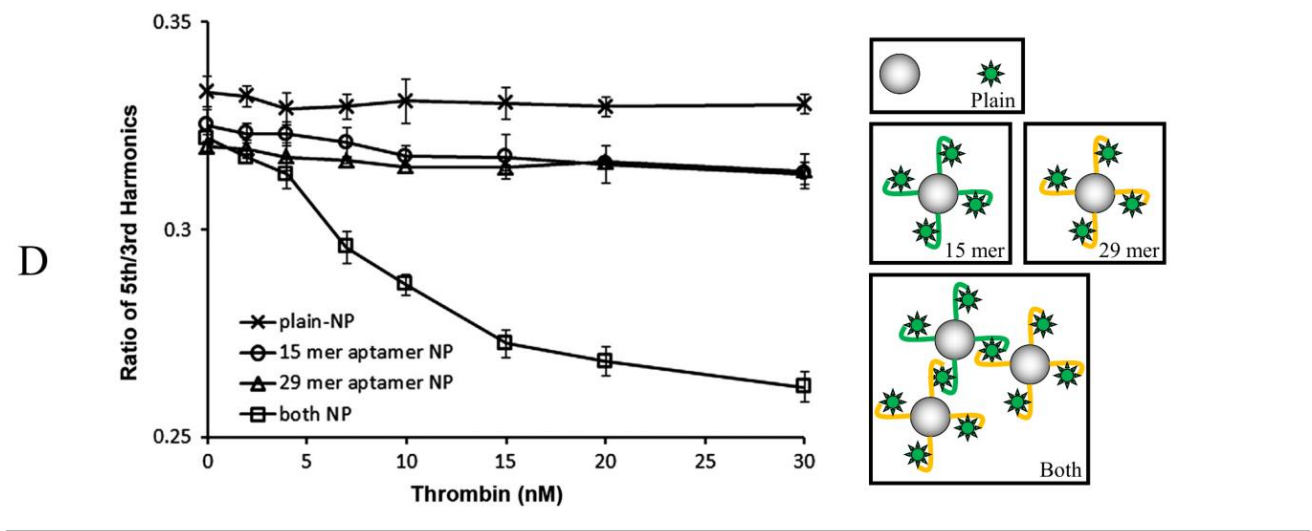
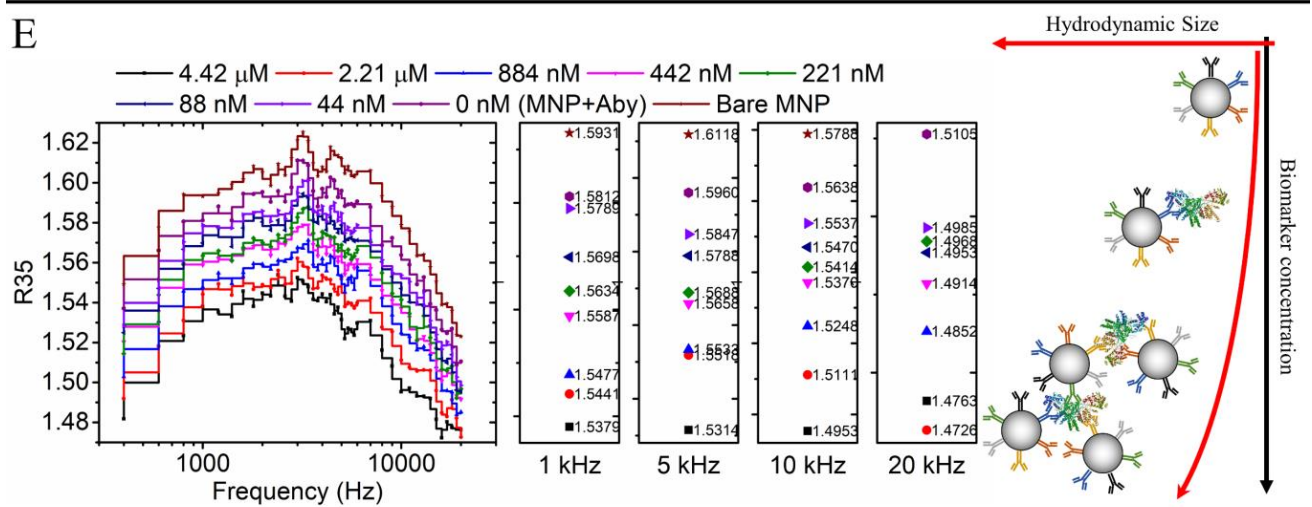

Figure 2. (A) Streptavidin and biotin functionalized SPION. As the quantity of streptavidin increases in the SPION liquid medium, the SPIONs are likely to form clusters. The dashed lines represent the hydrodynamic sizes of SPIONs due to the clustering. As the SPION clustering level increases, the hydrodynamic size increases, and the harmonic amplitude decreases. (B) Harmonic ratio of the $3^{rd}$ over the $5^{th}$ (R35) collected from six SPION samples added with different concentrations (quantities) of streptavidin, plotted as a function of driving field frequency. The inset figure summarizes the R35 under a drive field frequency of 10 kHz. (C) Left: bright-field TEM images of six SPION samples. Different SPION cluster models are given. Right: TEM images of SPIONs samples (15 µM, 7.5 µM, 3.8 µM, 75 nM, control (0 nM)) showing that the degree of particle clustering increases as the quantity (concentration) of streptavidin increases. (D) Harmonic ratio of the $5^{th}$ over the $3^{rd}$ (R53) from two populations of SPIONs, conjugated with 15 mer and 29 mer anti-thrombin aptamer, as a function of thrombin concentration. (E) Harmonic ratio of the $3^{rd}$ over the $5^{th}$ (R35) collected from nine SPION samples added with different concentrations (quantities) of H1N1 NP, plotted as a function of driving field frequency. Insets highlight the R35 measured at 1 kHz, 5 kHz, 10 kHz and 20 kHz, respectively. (A-C) reprinted with permission from [29], Copyright (2019) American Chemical Society. (D) reprinted with permission from [17], Copyright (2013) Elsevier. (E) reprinted with permission from [48], Copyright (2020) American Chemical Society.

It is already known that the binding events of target analytes affects the Brownian process and dynamic magnetic responses of SPIONs under external fields. While this detection mechanism yields detectable changes in harmonic amplitudes, phases, and harmonic ratios, larger changes in these metrics can be achieved by designing multiple ligands/antibodies/aptamers/proteins on SPIONs that could bind to hetero binding domains on the same biomarker, assembling the SPIONs into clusters or aggregates. This kind of detection mechanism greatly increased the hydrodynamic sizes of SPIONs, resulting in increased relaxation time, which can be sensed by the MPS system.

The nonspecific binding (NSB) between other molecules/chemicals to the SPIONs is a critical problem in the design of MPS-based immunoassays as the purpose of MPS bioassays is to measure only the binding events between the target analytes and the SPIONs, instead of the background. Some frequently used methods to prevent NSB are suggested herein as below: 1) adjust the pH of SPION liquid medium; 2) use protein blockers such as bovine serum albumin (BSA) to block spare binding sites on SPIONs; 3) use surfactants such as Tween 20 to disrupt hydrophobic interactions between target analytes and SPIONs; 4) adjust salt concentration in SPION liquid medium to prevent charges on the target analytes from interacting with charges on the SPIONs.

**2.2 MPS-based Cell Analysis**
SPIONs have been used for a wide variety of applications such as molecular and cell separation, drug delivery, hyperthermia, magnetic resonance imaging (MRI), and MPI, etc.[9,49–62] The need for quantitative methods to evaluate the SPION (or iron) content from contrast media solutions and biological matrixes is thus obvious. There are various tools reported to measure SPION (or iron) content such as spectrophotometric elemental analysis techniques (e.g., Perls' Prussian blue colorimetry), fluorophores/radionuclides labeled SPIONs, and R1 relaxometry.[63–66] Despite the extensive work done to quantify iron in an aqueous matrix, these techniques suffer from: inability to distinguish between exogenous SPIONs and endogenous iron, narrow detectable range, and nonlinearity.[36,63] Recently, MPS is reported as an alternative to measure iron content (or to quantify SPIONs) from aqueous matrix, where MPS utilizes the non-linear magnetic responses from SPIONs exposed to excitation magnetic field.

Loewa *et. al* reported the quantification of SPIONs uptaken in cells using MPS platform,[36] where they took the 3$^{rd}$ harmonic amplitude as a measure for the SPION content. To this end, the harmonic amplitudes of known SPION (iron) content under the same excitation field condition are recorded. As shown in Figure 3(A), the calibration curves are highly linear for all types of SPIONs. Two tumor cell lines, HeLa and Jurkat, were incubated with SPIONs of different concentrations, for 30 h, followed by harvesting ~$10^6$ cells for MPS measurements. The 3$^{rd}$ harmonic amplitudes from these harvested cells were recorded in Figure 3(B), indicating the SPIONs were up taken by cells. Using the calibration curves, the calculated iron content up taken by cells are given in Figure 3(C). Although this is a preliminary work without considering the impact of SPION aqueous matrix (viscosity, binding, aggregations, etc.) and the influence of size selective cellular uptake of SPIONs, this method provides the basis of a practical implementation of *in vivo* studies of SPION (iron) content from tissue samples. Besides the harmonic amplitude, the harmonic ratio R53 is also reported to be an indicator to reveal distinct changes in the magnetic behaviors of SPIONs in response to cellular uptake.[37]

Gräfe *et. al* established an *in vitro* testing system to investigate SPION transport across cellular layers as well as examining MPS for reliable SPION quantification.[38] A blood-brain barrier (BBB) representing human brain microvascular endothelial cells (HBMEC) model was chosen. Their experiment confirms the excellent suitability of MPS for sensitive SPION quantification at different stages of particles passing cellular layers, indicating a promising future method to pass SPIONs loaded drugs across BBB without disturbing its integrity. As a result, the drug loading and release efficiency can be monitored by MPS approach.

Fidler *et. al* reported MPS for *in vitro* cell vitality monitoring, where human mesenchymal stem cells (hMSCs) are labeled with SPIONs and tracked.[39] The hMSCs are a promising tool in regenerative medicine and it's able to repair damaged tissue. However, the tissue healing using hMSCs will only be possible if cells can be homed to their target and are still vital. This challenge calls for a long-term, non-invasive method to label and monitor the cells. Fidler *et. al* labeled hMSCs with SPIONs and the verification of SPION-labeled hMSCs by Prussian blue, TEM, and MPS, the light and TEM analysis is shown in Figure 3(D) & (E). For cell vitality assessment, the MPS spectra was continuously monitored during a cell degradation process initiated by adding Sodium Dodecyl (lauryl) Sulfate (SDS, which dissolves the cells). As shown in Figure 3(F), during the dissolution of the cells by SDS, the MPS harmonic amplitudes increase and larger changes are observed from higher harmonics. The increase of harmonic amplitudes can be attributed to a change in the Brownian relaxation, where, during the cell degradation process, cells are dissolved and SPIONs are released to a lower viscosity extracellular environment (or from bound to unbound status). This work demonstrated the feasibility of using SPIONs with MPS platform as reporters for cell homing and cell vitality, in an aqueous environment.

In summary, the efficient noninvasive techniques for tracking cells in cell-based therapies and diagnostics is crucial. SPIONs, with superior biocompatible, biodegradable and superparamagnetic nature, are much desired magnetic markers for labeling cells. Upon the interaction with biological materials such as cells, SPIONs undergo physicochemical changes that alters their dynamic magnetic responses and eventually the MPS spectra.[37] Currently, there are many established techniques including chromatographic and colorimetric iron assays and electron microscopy, which, are not feasible with living cells. MPS may aids in overcoming these limitations.

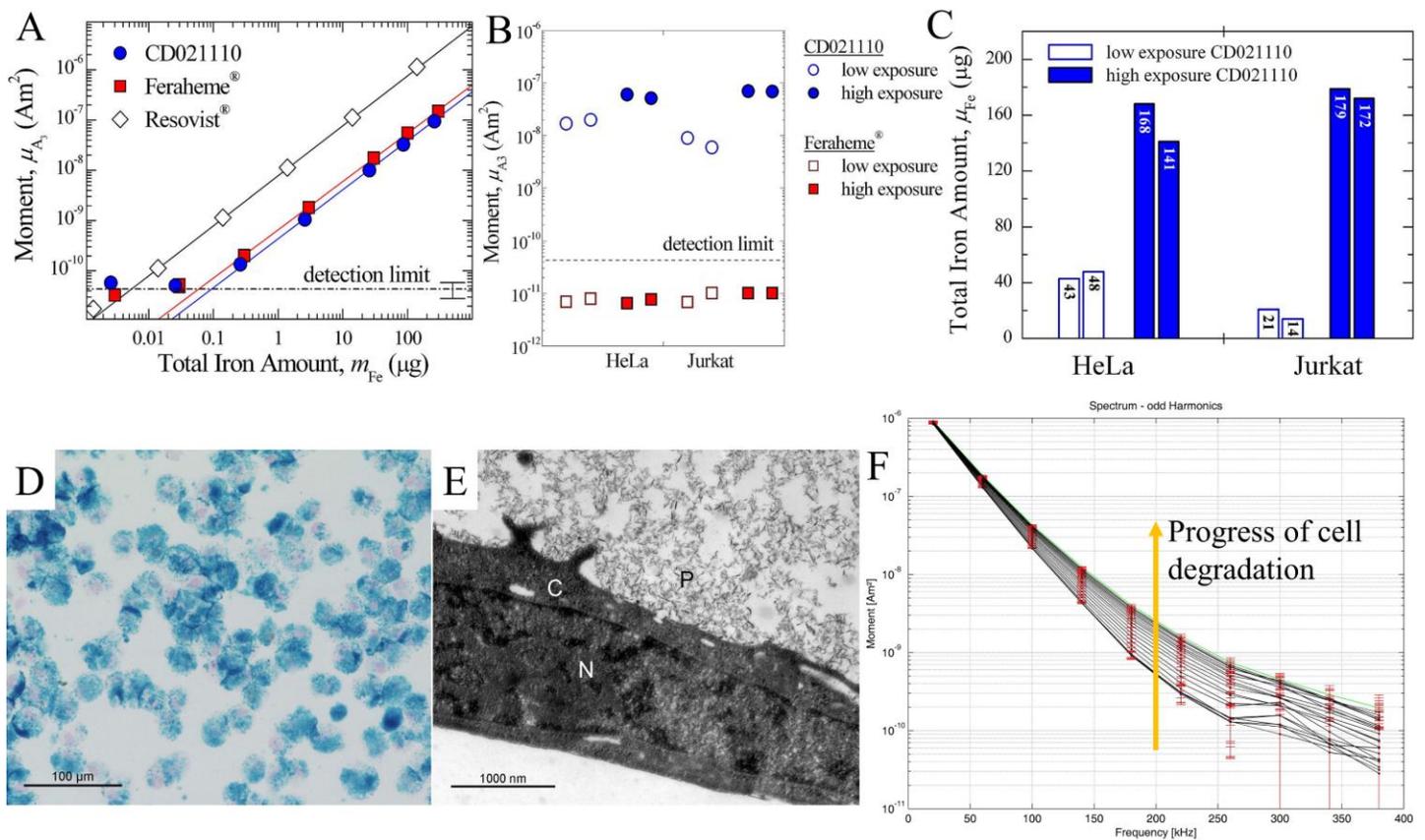

Figure 3. (A) Calibration curves of Feraheme, CD021110, and Resovist (a tracer most frequently used for MPI) SPIONs. (B) MPS signal (the 3$^{rd}$ harmonic amplitude) from harvested cells (HeLa and Jurkat) incubated with low and high concentrations of CD021110 and Feraheme SPIONs. (C) Cellular uptake of CD021110 SPIONs by HeLa and Jurkat cells quantified by MPS. The higher the dose of nanoparticles, the more particles have been resorbed by the cells. (D) Prussian blue staining verifies the presence of SPIONs, nuclei appear in light red. (E) TEM shows extracellular localization of SPIONs adhering to the cell's surface. C: cytoplasm; N: nucleus; P: SPIONs. (F) MPS spectra of degradation and degraded (gray/green line) stem cells measured at 20 kHz and 37 °C.

The degradation process was continuously monitored. (A-C) reprinted with permission from [36], Copyright (2012) IEEE; (D-F) reprinted with permission from [39], Copyright (2015) IEEE.

## 3. MPS as Auxiliary Tool

### 3.1 MPS for Viscosity and Temperature Monitoring

The Brownian relaxation process is affected by the viscosity of SPION liquid medium. Thus, this property has been applied for *in vivo* and *in vitro* viscosity monitoring using Brownian process-dominated SPIONs.[15,20–22,67,68] As shown in Figure 4(A), SPIONs are dispersed in different aqueous medium of varying viscosities. At a driving field frequency of 500 Hz, SPIONs from low viscosity medium show more higher harmonics and the slope of harmonic curve increases as viscosity increases. Because under low frequency excitation fields, magnetic moments of SPIONs can follow the field directions almost instantaneously via the Brownian relaxation. The only countering force is the friction force from high viscosity medium. Therefore, there is a phase lag between magnetic moments (due to the slow response of Brownian relaxation process) and the excitation field. The SPIONs do not reach equilibrium states before the excitation field changes direction. This relationship is reflected in the series of dynamic M-H curves in Figure 4(A). Using this property of Brownian relaxation process, Wu *et. al* proposed and demonstrated the feasibility of using MPS and SPIONs to measure human serum viscosity in real-time.[21] A standard calibration curve of $3^{rd}$ harmonic amplitude *vs.* viscosity was plotted, for estimating any aqueous mediums with unknown viscosities.

In addition, the temperature information of SPIONs' aqueous medium can also be extracted from their magnetic responses due to the fact that Brownian relaxation process is also modulated by the temperature (thermal fluctuation). Several groups have reported the *in vivo* temperature measurements using MPS and SPIONss, where the correlations between temperature and Brownian relaxation time are obtained by measuring the MPS spectra (magnetic responses) across a range of frequencies and temperatures. Calibration curves are plotted to subsequently estimate the temperatures.[25,69,70]

In short, the temperature and viscosity information of the aqueous medium can be separated from the magnetic responses of SPIONs. In recent years, there are several research groups reporting the methods of simultaneous MPI and temperature/viscosity mapping using multi-SPION contrasts. In these studies, MPI provides SPION distribution images with high temporal and spatial resolution and, meanwhile, the temperature/viscosity information is also separated from the MPS spectra. In 2016, Stehning *et. al* reported simultaneous MPI and temperature imaging using multi-color reconstruction approach.[40] The proof of principle for the experimental setup is shown in Figure 4(B). Two SPION samples (one supplied with tempered distilled water and the other is non-tempered as reference) were mounted on a plastic hose at a distance of 15 mm. The reconstructed SPION amounts from both samples with same size, $C_{REF}$ and $C_{TEMP}$, were equivalent. The temperature signal of the non-tempered reference SPION sample, $T_{REF}$, remained constant over the imaging experiment. The temperature signal of the tempered SPION sample, $T_{TEMP}$, declined as the circulating water temperature is lowered, as shown in Figure 4(C). Furthermore, the spatially resolved MPI and temperature mappings are depicted in Figure 4(D). The initial frames acquired at 37 °C are shown in upper row, while, the final frames acquired after the circulation of water when tempered sample is at 12 °C are shown in the lower row. The color-coded subtraction images Figure 4(D: [e] & [f]) represent the temperature map. This method may allow for real-time temperature monitoring in image-guided interventions, such as interstitial hyperthermia, following a direct injection of nanoparticles. Later Möddel *et. al* reported a viscosity quantification method using multi-contrast MPI.[20]

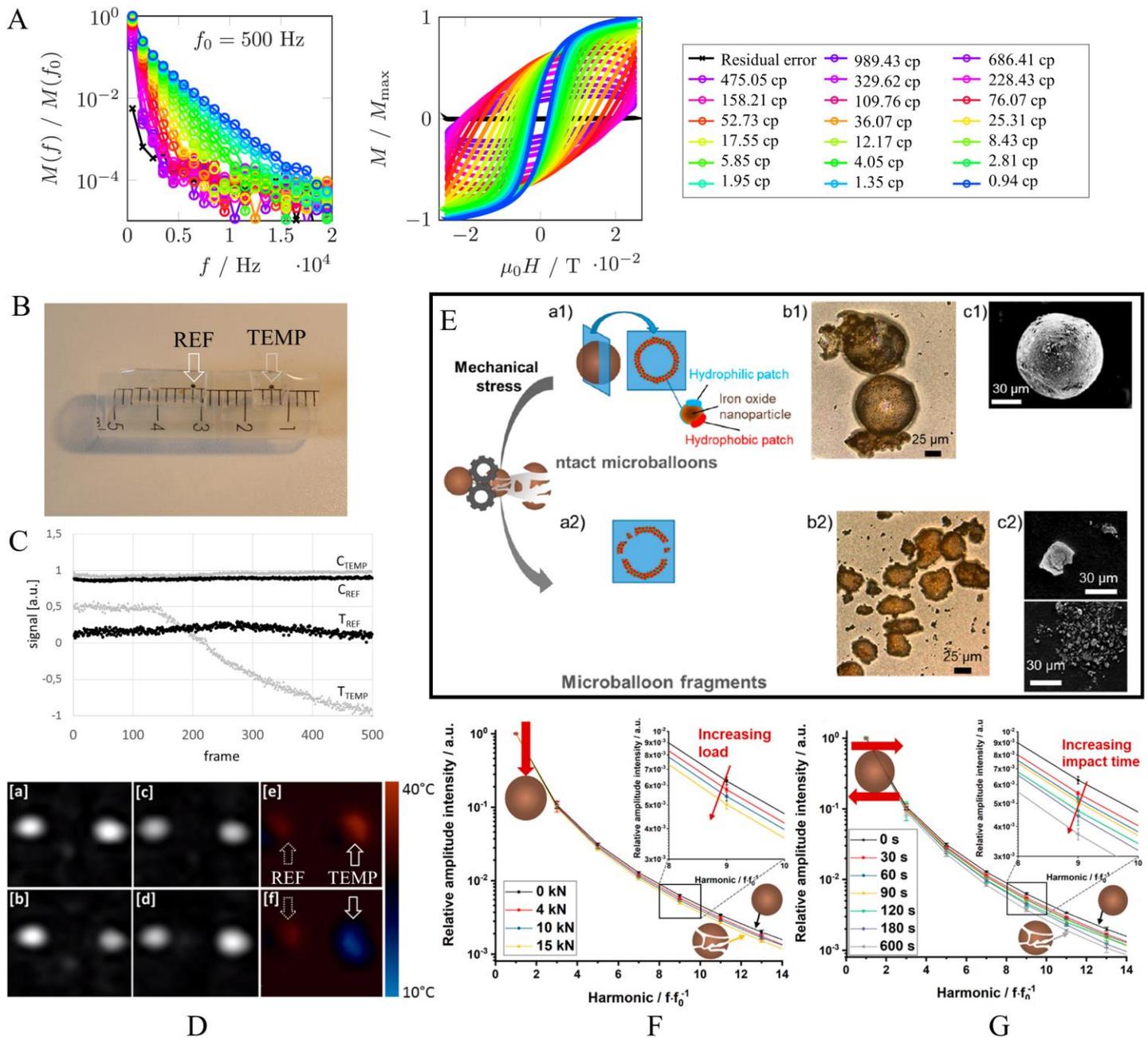

Figure 4. (A) Experimental results of MPS harmonic amplitudes *vs.* viscosity series acquired under driving field of 500 Hz and 25 mT. Black lines represent residual error R. Measurements performed at room temperature T=25 °C. (B) Plastic tube with two SPION samples. (C) Reconstructed MPI signals reflecting the amount of SPIONs ($C_{TEMP}$ and $C_{REF}$) and temperature information ($T_{TEMP}$ and $T_{ERF}$) of the tempered and reference probe. (D) Projection images of 3D reconstructed signal channels [a]-[d], and color-coded difference signal representing temperature [e], [f] acquired at the beginning [top row] and end [bottom row] of the experiment. (E) Hollow microballoons composed of SPIONs. Drop of the MPS signal curves of microballoons during the application of quasi-static compression (F) or of dynamic shear and impact forces in a ball mill-like setup (G) increases significantly with increasing load or milling time, respectively. (A) reprinted with permission from [15], Copyright (2019) American Chemical Society; (B-D) reprinted with permission from [40], Copyright (2016) Infinite Science Publishing. (E-G) reprinted with permission from [41], Copyright (2019) American Chemical Society.

### 3.2 MPS for Mechanical Force Monitoring

Wintzheimer *et. al* reported a novel application of MPS and SPIONs as sensors for mechanical stress.[41] The as-assembled SPIONs yield hollow microballoons (as shown in Figure 4(E)). This kind of structure is continuously fragmented when subjected to mechanical forces. By using MPS spectra, this structure deformation can be readily detected and, enables quantification of the applied mechanical forces in ball mills. Figure 4(F) & (G) depict the MPS harmonic amplitudes of the microballoons after the application of quasi-static compression or dynamic shear and impact forces, respectively. In both cases, the relative amplitude intensities drop with the increasing mechanical energy, which proves that the MPS spectra of these hollow microballoon supraparticles are clearly distinguishable in the course of mechanical forces. This type of mechanical force sensor does not necessarily have to be removed from the processed material for measurements, furthermore, its magnetic property allows easy removal after milling.

### 4. Other applications
### 4.1 MPS for SPION Characterization

As the sinusoidal magnetic fields excite SPIONs into their nonlinear saturation magnetization. The dynamic magnetic responses, represented by higher harmonics in frequency domain, provide information about the physical and magnetic properties of the SPIONs. It has been reported that the harmonic ratio R35 is inversely proportional to the saturation magnetization $M_s$ and magnetic core diameter D of the SPION.[15,42,67,71] The harmonic phase angle (lag) of magnetic moment to the excitation field carries the information of hydrodynamic size of SPION. Numerical simulations were carried out to reveal the correlation between harmonic ratio and $M_s$ as well as D. As shown in Figure 5(A: A1 & A2), a SPION system is assumed, with log-normal size distribution, effective anisotropy constant $K_{eff} = 1.8 \times 10^5$ erg/cm$^3$ ($1.8 \times 10^4$ J/m$^3$), $M_s$ and D are varied. The results show that for SPIONs with identical core sizes, a smaller harmonic ratio R35 corresponds to a higher $M_s$, while, for SPIONs with identical $M_s$, a smaller core size yields a larger harmonic ratio R35. This conclusion is further proved by experimentally measuring the harmonic ratios from four commercially avaible SPION samples, as shown in Figure 5(A: A3). For SMG30-I and -II samples, with identical magnetic core size, a smaller harmonic ratio R35 from SMG30-II indicates a higher $M_s$ over SMG30-I, which is further proved by the Vibrating Sample Magnetometer (VSM) results and high-angle annular dark-field scanning transmission electron microscopy energy-dispersive X-ray spectroscopy (HAADF–STEM–EDS) mapping images.[42] In addition, the 3$^{rd}$ harmonic phase lag to excitation field recorded in Figure 5(A: A4) shows that SMG30-I and -II SPIONs have similar hydrodynamic size, and are larger than SHP25. On the other hand, for the multicore MACS SPIONs, the SPIONss are embedded in a polymer and Néel process is dominating, which is a different scenario compared to the single core SPIONs (SHP25, SMG30-I and -II).

In addition to the magnetic and physical properties of SPIONs that affect the MPS spectra (harmonic amplitudes and phases), the excitation fields should also be taken into consideration. Draack *et. al* reported the field-dependent MPS spectra by measuring SPIONs under different exaction field amplitudes and frequencies.[15] The field-dependent harmonic amplitudes are measured at a constant excitation frequency, f=1 kHz, with varying amplitudes from 5 mT to 25 mT, as shown in Figure 5(A: A5). The dynamic M-H curves are plotted in Figure 5(A: A6). A higher magnetic field strength results in more saturated magnetizations and more pronounced hysteresis of the magnetization loops.

**4.2 MPS for Magnetic Fingerprint**

Based on the aforementioned phenomenon, each type of SPIONs with different magnetic core size D, hydrodynamic size, saturation magnetization $M_s$, shows unique MPS spectra, which could be used as magnetic fingerprints for product tracking and identification. Müssig *et. al* reported the specially engineered magnetic supraparticles that exhibit unique magnetic signature in MPS, proving its potential for marker applications.[43] In their work, the supraparticles are formed by assembling SPIONs with $SiO_2$ nanoparticles with different ratios, as shown in Figure 5(B: B3-B6). They prepared different kinds of supraparticles by mixing SPIONs (~10 nm) and $SiO_2$ nanoparticles (~20 nm) in weight ratios of 1:0, 1:1, 1:2, and 1:4. The M-H curves measured by VSM are plotted in Figure 5(B: B1), yet, the differences with respect to hysteresis loops can hardly be quantified. However, clearly distinguishable MPS spectra are observed in Figure 5(B: B2). In contrast to the VSM, MPS spectra of various supraparticles are easily quantified by determining the amplitudes of higher harmonics. In addition to mixing SPIONs with non-magnetic $SiO_2$ nanoparticles, changing the interparticle distance by surface modification of SPIONs with OCTEO can also modify MPS spectra (as shown in Figure 5(B: B1-B2), the "$Fe_3O_4$ part/full OCTEO-modified" curves). This work shows such spraparticles along with MPS could enable easy and versatile tracking and identification for products. By mixing different amount of $SiO_2$ nanoparticles with SPIONs or structurally adjusting the SPIONs, it is possible to change the magnetic properties (MPS spectra) of supraparticles and generate countless unique magnetic fingerprints.

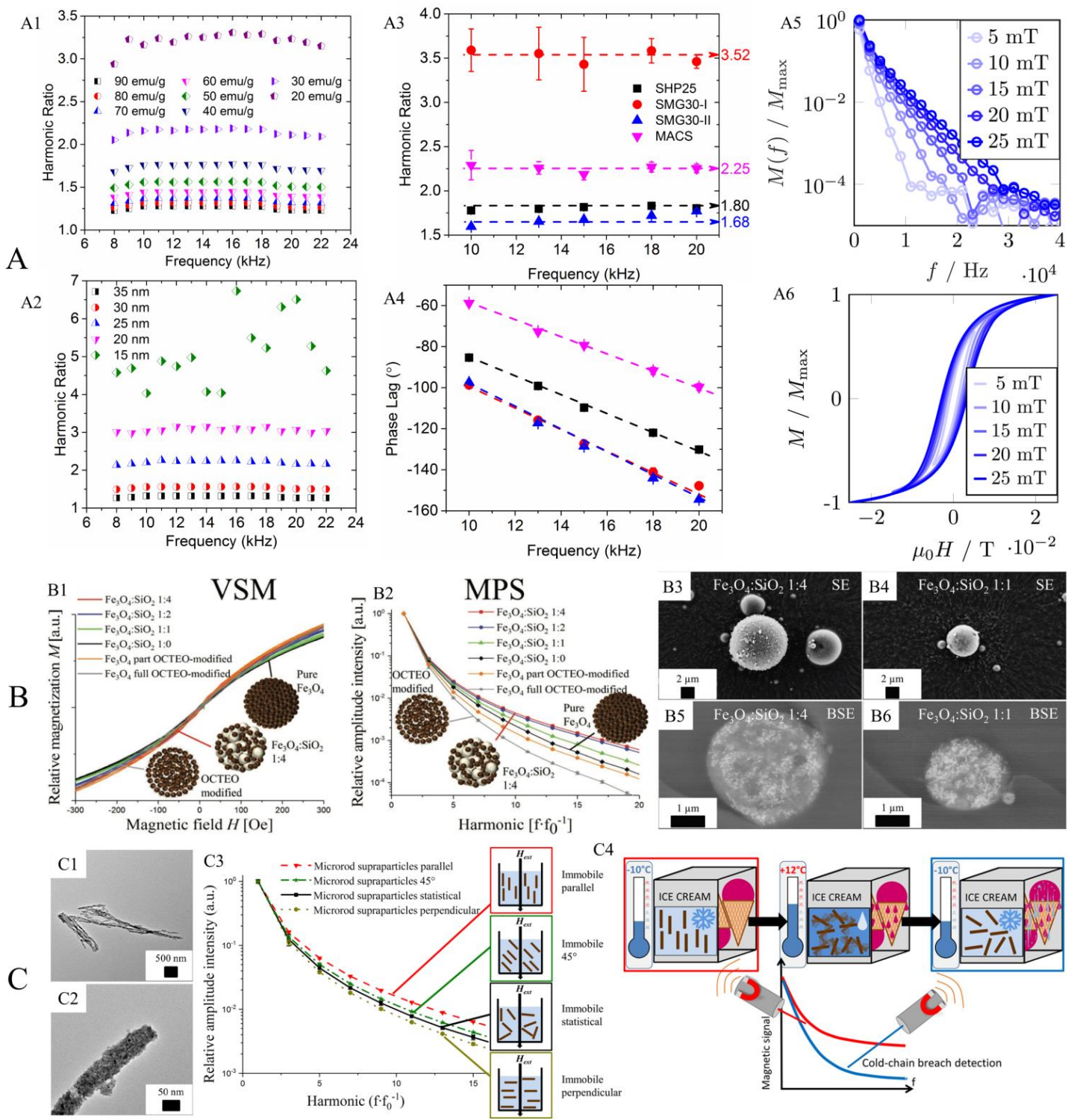

Figure 5. (A) MPS for SPION characterizations. A1: Simulated harmonic ratios R35 as a function of excitation field frequency when D=30 nm, $M_s$ of SPIONs varied from 20 to 90 emu/g. A2: Simulated harmonic ratios R35 for different SPION core sizes with identical $M_s$ = 50 emu/g as a function of excitation field frequency. A3: Measured harmonic ratios R35 from SHP25, SMG30-I, -II, and MACS SPION samples as function of excitation field frequency. A4: Measured phase lags at the 3$^{rd}$ harmonics as function of excitation field frequency. A5: Measured field-dependent harmonic amplitudes at f=1 kHz with varying amplitudes. A6: Measured field-

dependent dynamic M-H curves acquired at f=1 kHz with varying amplitudes. (B) B1: Normalized VSM magnetization curves of different supraparticles. B2: Relative amplitude intensity as a function of the higher harmonics of the same samples measured via MPS. B3-B4: SEM images of supraparticles consisting of SPIONs and $SiO_2$ nanoparticles in weight ratios of 1:4 and 1:1. B5-B6: Backscattered electron detector (BSE) images visualizing the material contrast between the bright SPIONs and darker $SiO_2$ nanoparticles in the supraparticles with weight ratios of 1:4 and 1:1. (C) C1-C2: TEM images of anisotropic microrod supraparticle morphologies. C3: MPS spectra of immobilized anisotropic microrods differently oriented with respect to the external excitation field. C4: Schematic depiction of ice cream packages equipped with a microrod-supraparticle-based temperature indicator. (A1-A4) reprinted with permission from [42], Copyright (2017) Wiley; (A5-A6) reprinted with permission from [15], Copyright (2019) American Chemical Society; (B) reprinted with permission from [43], Copyright (2019) Wiley; (C) reprinted with permission from [44], Copyright (2019) American Chemical Society.

Microrods assembled from SPIONs have also been reported as magnetic fingerprints, since the orientation-dependent MPS spectra variations of anisotropic microrods are identifiable, as shown in Figure 5(C: C1-C3).[44] When the microrod supraparticles are oriented in parallel and perpendicular with respect to the excitation field, the MPS spectra with highest and lowest intensities are observed, respectively. Figure 5(C: C4) schematically depicted the ice cream packages equipped with a microrod-supraparticle-based temperature indicator. The initially aligned and frozen microrod supraparticles become mobile when an undesired temperature increase melts the frozen matrix. The refreezing of the matrix immobilizes the microrod supraparticles in a different alignment and thus makes a potential quality loss detectable by their respective MPS curves. This unique property of anisotropic supraparticles enables the detection of cold-chain breaches (e.g., during delivery of a product that requires to be cooled all the time), simply by recording the initial and final MPS spectra of microrod supraparticles integrated into the container of a frozen product.

**4.3 MPS-based Blood Analysis**

Determining the patient's total blood volume is an essential topic in clinical routine. There are a variety of methods reported but due to the extend volume and dispersion throughout the body, a direct measurement seems unfeasible. A promising approach to determine total circulating blood volume is based on the dilution measurement of applied tracers. Franziska *et. al* reported a MPS method along with SPION tracers for this purpose.[45] They used FeraSpin™ R SPIONs with hydrodynamic diameter of 60 nm and blood half-life in rat of 15 min. However, for a prolonged blood half-life, another product FeraSpin™ XS with hydrodynamic diameter of 15 nm and blood half-life in rat of 30 min is also recommended. This method consists of three major steps: First, estimate the undiluted SPION tracer and choose a specific volume of injection, second, the SPION sample injected into the subject, third, a small amount of blood is draw after a mixing time. From the measured concentration, the total circulating blood volume is calculated.

Another work reported by Hafsa *et. al* proposing the evaluation of blood clot progression using SPION tracers and MPS.[16] By measuring the SPIONs' relaxation time in the clot vicinity, the clot age, clot stiffness, and the amount of SPIONs bound to the clot can be estimated. It is reported that during the clot formation, SPIONs become trapped in the clot mesh, which, restricts their Brownian relaxation process, causing an increase of their relaxation time.[46] In this work, the SPIONs are surface functionalized with ATP15 and ATP29 aptamers that bind to hetero epitopes on thrombin. SPIONs bound to thrombus are identified by their increased relaxation time using MPS. The harmonic ratio R35 is reported as metric for the Brownian process, which in turn, reflects the bound state of SPIONs.

Table 1. Summary on MPS-based Applications

| Applications | SPION ($d_{core}$) | Driving Field | Assay time | Analyte | Matrix | Ref. |
|---|---|---|---|---|---|---|
| **Immunoassay** | $d_{core}$~30 nm | 17 mT, 500 Hz – 20 kHz<br>1.7 mT, 10 Hz | 10 s | Streptavidin | PBS buffer | [29] |
|  | $d_{core}$~50 nm | 10 mT, 290 Hz – 2110 Hz | 5 s | Thrombin | PBS buffer | [17] |
|  | $d_{core}$~30 nm | 17 mT, 500 Hz – 20 kHz<br>1.7 mT, 10 Hz | 10 s | H1N1 NP | PBS buffer | [48] |
| **Monitor cellular uptake** | Feraheme ($d_{core}$~5 nm),<br>CD021110 ($d_{core}$~4 nm) | 25 mT, 25 kHz | - | Tumor cell lines (HeLa and Jurkat) | Cell culture medium | [36] |
|  | $d_{core}$~7 nm | 25 mT, 25 kHz | - | THP-1 cells | Cell culture medium | [37] |
| **Cell vitality assessment** | - | 20 mT, 20 kHz | - | hMSCs | Cell culture medium | [39] |
| **Monitor particle uptake and transport at cellular barriers** | Hydrodynamic size ~140 nm | 25 mT, 25 kHz | - | HBMEC | Cell culture medium | [38] |
| **Viscosity monitoring** | $CoFe_2O_4$, $d_{core}$=15.5 nm | 25 mT, 500 Hz | - | - | Water-glycerol mixtures | [15] |
|  | $d_{core}$=25 nm | 10 mT, 15 kHz<br>1 mT, 50 Hz | 1.5 min | - | Human serum | [21] |
| **Simultaneous MPI and temperature mapping** | - | - | - | - | - | [40] |

| Mechanical force monitoring | ~40 μm hollow microballoon supraparticle assembled from SPIONs | 30 mT, 20.1 kHz | - | - | - | 41 |
|---|---|---|---|---|---|---|
| SPION characterization | SHP25 ($d_{core}$~25 nm), SMG30-I ($d_{core}$~30 nm) SMG30-II ($d_{core}$~30 nm) MACS (multi-core, hydrodynamic size ~50 nm) | 10 mT, 8-22 kHz 1 mT, 50 Hz | - | - | PBS buffer | 42 |
| | $CoFe_2O_4$, $d_{core}$=15.5 nm | 5-25 mT, 1 kHz | - | - | - | 15 |
| Magnetic fingerprint | SPIONs of ~10 nm and $SiO_2$ of ~10 nm are mixed and spray dried | 30 mT, 20 kHz | - | - | - | 43 |
| | Microrods assembled from SPIONs of ~10 nm. | 30 mT, 20.1 kHz | - | - | - | 44 |
| Determine the total circulating blood volume | FeraSpin<sup>TM</sup> R (nanoPET Pharma GmbH, Berlin, Germany) | 20 mT, 20 kHz | - | - | Rat blood | 45 |
| Evaluate blood clot progression | $d_{core}$~50 nm, hydrodynamic size 116 nm | 10 mT, 400 – 1600 Hz | - | - | Pig blood | 16 |
| | $d_{core}$~25 nm, hydrodynamic size 40 nm | 10 mT, 25 kHz | 30 s | - | Human blood | 46 |

## 5. Conclusion

In this review, we summarized the recent advances in MPS-based applications and shown in Table 1. For biological and biomedical assays, by surface functionalizing suitable ligands/antibodies/aptamers/proteins on SPIONs that could specifically bind to target analytes, SPIONs are explored as magnetic markers and their dynamic magnetic responses are monitored by MPS. The binding events of target analytes affect the Brownian process and dynamic magnetic responses of SPIONs and cause detectable changes in harmonic amplitudes, phases, and harmonic ratios. Due to the negligible magnetic background from biological samples, highly sensitive detection could be achieved by using SPIONs as biomarkers. This detection mechanism based on a MPS platform allows simple, fast, inexpensive, and high sensitivity bioassays. In the areas of cell labeling, tracking and cell-based theragnostics (therapy and diagnostics), MPS provides a quantitative method to evaluate the SPION (or iron) content from contrast biological matrixes, and even in living cells. Upon the interaction with biological materials such as cells, SPIONs undergo physicochemical changes that alter their MPS spectra. In addition to these stand-alone applications, MPS has also developed into a mature auxiliary tool supporting the magnetic imaging and hyperthermia by providing real-time, high spatial and temporal resolution viscosity/temperature mapping. Other novel applications based on MPS such as SPION characterizations, magnetic fingerprints, and blood analysis are also reviewed.

In summary, many good and novel works have been reported in recent years based on MPS platform and this process is still ongoing. With the unique nonlinear magnetic responses of SPIONs, facile surface chemical modifications, flexible structural design of SPION assemblies (supraparticles) with novel magnetic/physical properties, MPS has found its position in more and more new areas.

## ASSOCIATED CONTENT
### ORCID
Kai Wu: 0000-0002-9444-6112

Diqing Su: 0000-0002-5790-8744

Renata Saha: 0000-0002-0389-0083

Jinming Liu: 0000-0002-4313-5816

Jian-Ping Wang: 0000-0003-2815-6624

### Notes
The authors declare no conflict of interest.

### ACKNOWLEDGMENTS

This study was financially supported by the Institute of Engineering in Medicine of the University of Minnesota through FY18 IEM Seed Grant Funding Program, the Distinguished McKnight University Professorship, the Centennial Chair Professorship, and the Robert F Hartmann Endowed Chair from the University of Minnesota.

TOC

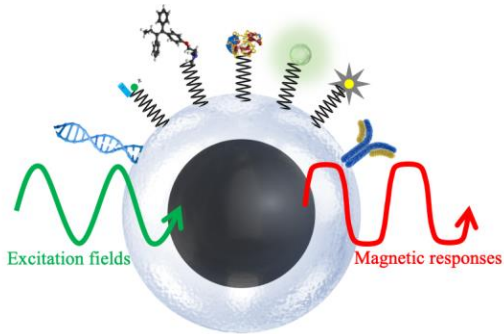